\documentclass[twocolumn,superscriptaddress,aps,pre,amssymb,amsfonts,amsmath]{revtex4-1}
\usepackage{bm}
\usepackage{graphicx}
\usepackage{xcolor}

\definecolor{darkblue}{rgb}{0,0,0.6}
\usepackage[colorlinks, linkcolor=darkblue, citecolor=darkblue, urlcolor=darkblue]{hyperref}

\begin{document}

\title{Is glass a state of matter?}

\author{Benjamin Guiselin}

\affiliation{ENSL, CNRS, Laboratoire de physique, Université de Lyon, F-69342 Lyon, France}

\author{Gilles Tarjus}
\affiliation{LPTMC, CNRS-UMR 7600, Sorbonne Universit\'e, 4 Pl. Jussieu, F-75005 Paris, France}

\author{Ludovic Berthier}

\affiliation{Laboratoire Charles Coulomb (L2C), Universit\'e de Montpellier, CNRS, 34095 Montpellier, France}

\affiliation{Yusuf Hamied Department of Chemistry, University of Cambridge, Lensfield Road, Cambridge CB2 1EW, United Kingdom}

\date{\today}

\begin{abstract}
Glass is everywhere. We use and are surrounded by glass objects which make tangible the reality of glass as a distinct state of matter. Yet, glass as we know it is usually obtained by cooling a liquid sufficiently rapidly below its melting point to avoid crystallisation. The viscosity of this supercooled liquid increases by many orders of magnitude upon cooling, until the liquid becomes essentially arrested on experimental timescales below the ``glass transition'' temperature. From a structural viewpoint, the obtained glass still very much resembles the disordered liquid, but from a mechanical viewpoint, it is as rigid as an ordered crystal. Does glass qualify as a separate state of matter? We provide a pedagogical perspective on this question using basic statistical mechanical concepts. We recall the definitions of states of matter and of phase transitions between them. We review recent theoretical results suggesting why and how an ``ideal glass'' can indeed be defined as a separate equilibrium state of matter. We discuss recent success of computer simulations trying to analyse this glass state. We close with some experimental perspectives.
\end{abstract}

\maketitle

\section{What is a glass?}

\label{sec:intro}

\begin{figure}
\includegraphics[width=\columnwidth]{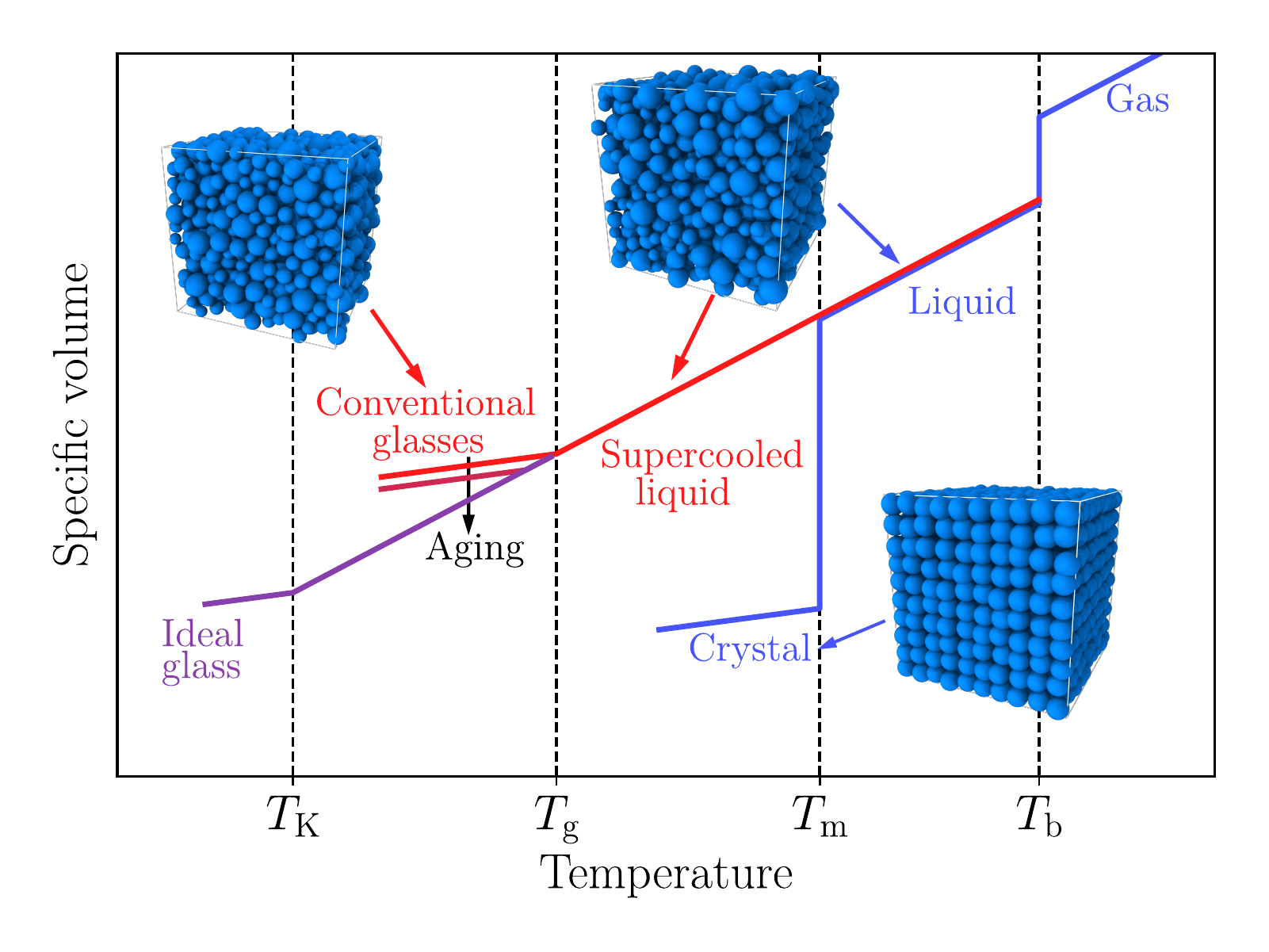}
\caption{{\bf Phases of matter.} A gas transforms into a liquid below the boiling temperature $T_\mathrm{b}$ through an equilibrium first-order phase transition. A liquid crystallises below the melting temperature $T_\mathrm{m}$ through another first-order phase transition. A metastable supercooled liquid can persist below $T_\mathrm{m}$ and gradually transforms into a solid at the glass transition temperature $T_\mathrm{g}$. Below $T_\mathrm{g}$, the glass is frozen in a rigid disordered structure which depends on the cooling rate. This is the common glass, which slowly ages towards equilibrium. An equilibrium ``ideal glass'' state would be obtained across a first-order phase transition at $T_\mathrm{K}$ provided thermal equilibrium could be maintained down to $T_\mathrm{K}$.}  
\label{fig:glass}
\end{figure}

We possess an intuitive notion of what a state of matter is. For a pure substance, there are usually three such states: gas, liquid, and crystal~\cite{tabor1991gases}. The gas is found at high temperatures and transforms into a liquid upon cooling. For an infinitely slow cooling process, where thermal equilibrium is maintained all along, condensation at the boiling point is a first-order equilibrium phase transition associated with a discontinuous jump in various thermodynamic quantities, such as the specific volume shown in Fig.~\ref{fig:glass} (blue curve). The liquid itself transforms abruptly into a crystalline solid when cooled below the melting temperature $T_\mathrm{m}$ through another first-order equilibrium phase transition. These states of matter are characterised and distinguished by their macroscopic properties. For example, the gas occupies the entire available volume, unlike the crystalline solid which retains its shape. 

Glass is absent from this description because we have so far focused on equilibrium properties. If a liquid is cooled quickly enough below $T_\mathrm{m}$ (red curve in Fig.~\ref{fig:glass}), crystallisation can be avoided and the system becomes a supercooled liquid~\cite{cavagna2009supercooled}. This supercooled state can exist over very long times because the formation of a crystal is very slow for a large number of substances. In this supercooled regime, the system appears for all practical purposes at thermal equilibrium although it is metastable with respect to the crystal. For instance, its physical properties do not depend on time and can easily be reproduced in independent experiments.

While the structure of the supercooled liquid changes very little with decreasing the temperature, its transport properties, such as viscosity or diffusivity, change by many orders of magnitude over a small temperature range. At the microscopic scale, molecular motion slows down dramatically and becomes so slow that very little motion can be detected over timescales of several hundreds of seconds when the temperature falls below the glass transition temperature $T_\mathrm{g}$. Below $T_\mathrm{g}$, the supercooled liquid is out of equilibrium, and molecules stop diffusing over experimentally accessible timescales~\cite{berthier2011theoretical}. This nearly-arrested liquid behaves  macroscopically like a solid even though its microscopic structure still resembles the one of a completely disordered liquid. The ``glass transition'' temperature at $T_\mathrm{g}$ is not a well-defined temperature and depends on the protocol used to cool the liquid. As a consequence, the glass that we use in our everyday life is a non-equilibrium material with physical properties that depend on its preparation history, see Fig.~\ref{fig:glass}, and may vary with time as the system slowly ages towards equilibrium. Therefore, a conventional glass is not a state of matter, as it is not an equilibrium material with crisply defined and fully reproducible thermodynamic properties. 

This is of course not the end of our story. The idea that a glass state of matter can exist in equilibrium conditions has a long history, dating back at least to the 
work of Kauzmann in 1948~\cite{kauzmann1948nature}. Kauzmann collected calorimetric data measured at equilibrium above $T_\mathrm{g}$ for a number of molecular liquids, and by extrapolation below $T_\mathrm{g}$ noted the possibility that a thermodynamic phase transition could take place at a temperature $T_\mathrm{K} < T_\mathrm{g}$, now called the Kauzmann temperature. Although Kauzmann himself favoured an interpretation in terms of a ``pseudocritical point'' marking the limit of stability of the liquid with respect to crystallization, his observation raised the prospect that the system below $T_\mathrm{K}$ could be in an equilibrium glass state. We shall call this state an ``ideal glass'', to make the distinction with conventional non-equilibrium glasses very clear. The Kauzmann temperature and the equilibrium ideal glass phase are shown in Fig.~\ref{fig:glass}. 

The hypothesis that an equilibrium phase transition at $T_\mathrm{K}$ underlies the formation of conventional glasses has given rise to a large litterature~\cite{berthier2011theoretical}. Our understanding of its nature has significantly evolved over the years since the early work of Gibbs and diMarzio who first described a simple lattice model with a thermodynamic liquid-glass transition~\cite{gibbs1958nature}. Here, we offer a pedagogical discussion of the ideal glass and the equilibrium glass transition relying on recent theoretical and numerical developments. Our aim is to make this discussion accessible to anyone with an undergraduate background in statistical mechanics at the level of well-known textbooks~\cite{landau2013statistical, callen1998thermodynamics, sethna2021statistical}.

In Sec.~\ref{sec:phase}, we review the concepts of states of matter and phase transitions to precisely define the question addressed in our title. In Sec.~\ref{sec:MF}, we summarise the mean-field theory of the glass transition and explain why, in this limit, glass does correspond to a state of matter. In Sec.~\ref{sec:numerics}, we present numerical results of glass-forming liquids which are consistent with the existence of an underlying liquid-to-glass 
equilibrium phase transition. We conclude in Sec.~\ref{sec:discussion}.

\section{What is a state of matter?} 

\label{sec:phase}

Following the conventional Landau approach to phase transitions~\cite{landau2013statistical} and without delving into all the subtleties, a state or a phase of matter is commonly defined as a homogeneous region of space in which all intensive thermodynamic variables are well-defined and uniform. We distinguish among these variables the control parameters (temperature, pressure, magnetic field, etc.) which can be imposed by an external operator, and others, which are densities associated with extensive quantities (specific mass, magnetisation, massic energy, etc.). The order parameter is one of these variables, but it plays a special role since it takes different values in the different phases, and its behaviour makes it possible to distinguish between them.

The different phases are usually represented by domains in a phase diagram, which is the space of control parameters. These domains are limited by phase transition lines through which the order parameter can vary continuously (second-order phase transitions) or discontinuously (first-order phase transitions). To determine the domain of stability of the different phases, Landau proposed a systematic method based on statistical mechanics~\cite{landau2013statistical}, which now corresponds to the starting point of most introductory courses on phase transitions~\cite{callen1998thermodynamics}. Having identified a good order parameter, one has to express the equilibrium free energy of the system as a function of this order parameter for fixed values of the control parameters (for example the temperature $T$), and then determine its absolute minimum for each value of $T$. The computed values of the order parameter and of the free energy at this minimum represent the value that would be measured in a macroscopic system at thermal equilibrium.

Some phase transitions are associated with a spontaneous symmetry breaking, corresponding to a change in the degree of symmetry of the system~\cite{goldenfeld2018lectures}. The variation of the order parameter from one phase to another reflects this loss of symmetry. For example, in the case of the transition between paramagnetic and ferromagnetic phases, the order parameter is the magnetisation vector which has a non-zero value in the ferromagnetic phase. The magnetisation then favours a single direction of space and thus breaks rotational invariance. Similarly, the crystallisation of a liquid into an ordered periodic lattice decreases the degree of symmetry of the system. Some phase transitions are not associated with a symmetry breaking, such as the liquid-gas transition.

To describe glass as an equilibrium phase along similar lines, a legitimate objection emerges immediately. Glass as we know it refers to a liquid that has left thermal equilibrium at $T_\mathrm{g}$ and can no longer explore its configuration space ergodically. This means that an equilibrium free energy cannot even be defined for a conventional glass and the above discussion appears irrelevant. Therefore, the questions we wish to address are: assuming thermal equilibrium can be maintained at arbitrary low temperatures and configuration space can be explored in equilibrium conditions, does the free energy show a transition from a liquid to a glass state, what is the order parameter that would then reveal the corresponding phase transition, and what is the corresponding symmetry breaking?

\section{Landau theory of the glass transition in the mean-field limit}

\label{sec:MF}

A series of works initiated in 2012 has led to the complete derivation of an analytic solution of the glass transition for simple models of glass-forming liquids which becomes mathematically exact in the limit of a large number of space dimensions, $d \to \infty$~\cite{parisi2020theory}. This non-physical mathematical limit makes the statistical mechanics of supercooled liquids exactly solvable. Such a situation is very common in statistical mechanics where it is often found that mean-field approximate solutions in finite dimensions become exact when $d$ becomes large~\cite{callen1998thermodynamics}. In large dimensions, the interactions between a given particle and its neighbours can be replaced by an average force which can be self-consistently determined. Despite this immense simplification, the calculation remains mathematically involved because it requires the explicit description of the aperiodic structure characterising the glass phase. The results found in large dimensions confirm a long series of earlier results which hinted at the nature of the glass transition when using mean-field approximations~\cite{singh1985hard, kirkpatrick1987pspin, biroli2001lattice}. 

It is possible to outline the main results obtained in this framework following the Landau approach described in Sec.~\ref{sec:phase}. To this end, the first step is to identify the good order parameter, $Q$, which can distinguish between the equilibrium liquid and glass phases, and then to express the free energy of the system as a function of $Q$, $V(Q)$. In this context, a suitable choice for $Q$ is what is called the overlap and for the associated $V(Q)$ the Franz-Parisi potential~\cite{franz1997phase}.

What indeed is a good order parameter $Q$? Clearly, neither the average density nor its spatial fluctuations can be used because if the glass looks like an arrested liquid then its density, $\rho$, its pair correlation function $g(r)$ and its static structure factor $S(k)$ also look similar. Therefore, contrary to the description of the liquid state where they play a key role~\cite{hansen2013theory}, density fluctuations are uninteresting spectators of the glass transition. 

The correct answer originates from another physics problem where glassy behaviour is also found, namely, spin glasses~\cite{mezard1987spin, castellani2005spin}. Although spin glass physics has a much shorter history than structural glass physics, the formulation of statistical mechanics models for spin glasses and a deep analytic understanding of their properties went comparatively much faster. Spin glass models describe spins with interactions mediated by random coupling constants, which makes it possible to describe the properties of dilute magnetic alloys~\cite{edwards1975theory}. The approriate order parameter to distinguish the spin glass and paramagnetic phases is the overlap $Q$ between magnetisation profiles of two independent copies (usually called ``replicas'' in the field of disordered systems) of the same system~\cite{edwards1975theory}. For Ising spins, this is simply expressed as $Q = (1/N)\sum_{i=1}^N S_i^{(1)} S_i^{(2)}$, where $S_i^{(1)}=\pm 1$ is the value of spin $i=1,\dots, N$ in one copy and $S_i^{(2)}$ the value of the same spin in the second copy. Clearly, the overlap $Q \simeq 1$ if copies 1 and 2 are highly correlated (spin glass), whereas for uncorrelated configurations $Q \simeq 0$ (paramagnet). The overlap $Q$ then quantifies the degree of similarity of the local degrees of freedom (here, the spins) in two configurations of the system drawn from the equilibrium Boltzmann distribution.

\begin{figure}
\includegraphics[width=0.95\columnwidth]{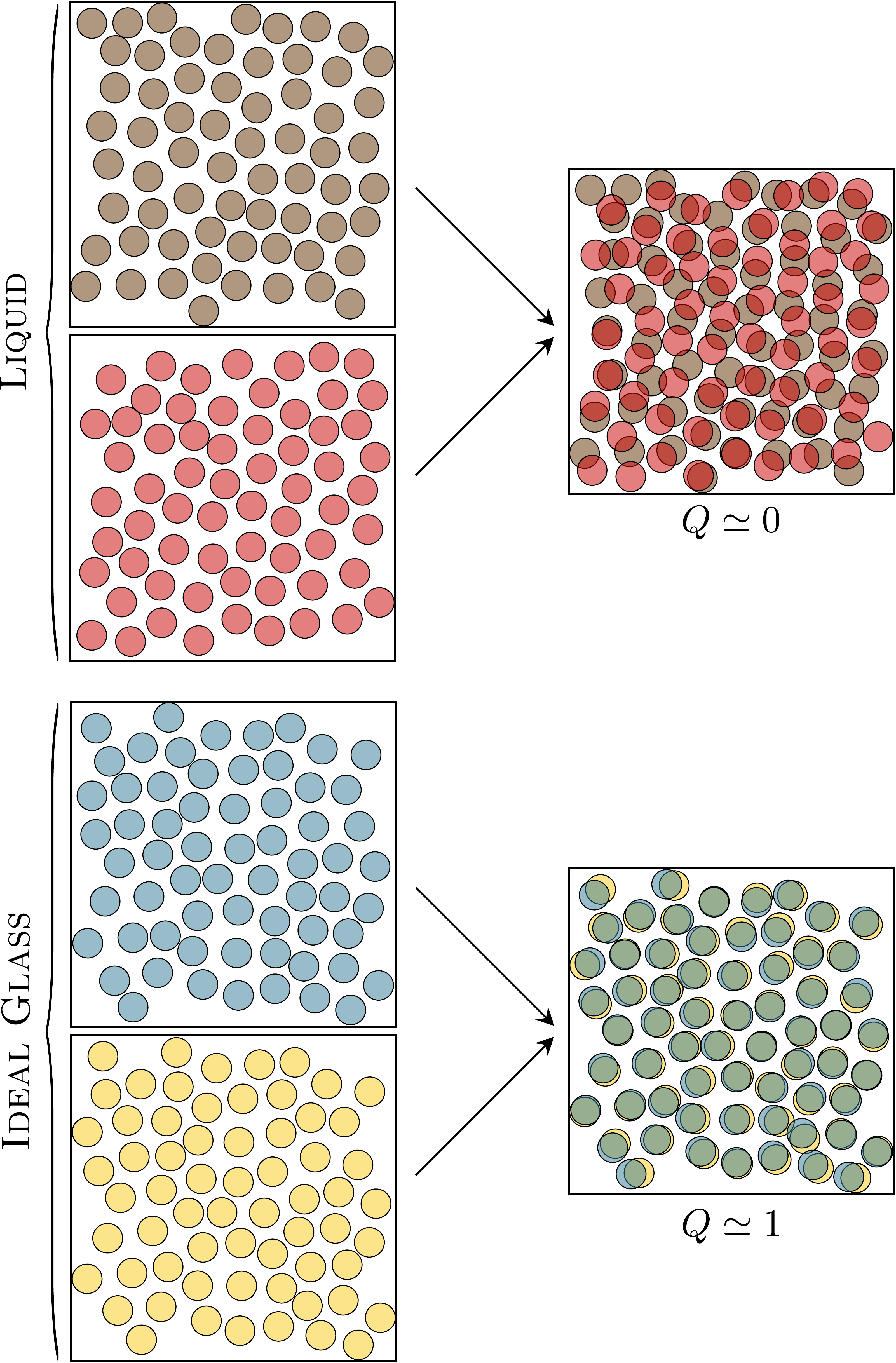}
\caption{{\bf Overlap order parameter $Q$ of the liquid-glass transition.} It quantifies the degree of similarity between the density profiles of two independent equilibrium configurations (here obtained from numerical simulations of a two-dimensional glass-forming liquid~\cite{guiselin2022statistical}). For $T>T_\mathrm{K}$, the liquid explores a large number of equilibrium structures and $Q \simeq 0$. Instead, in the equilibrium glass for $T<T_\mathrm{K}$, all equilibrium configurations are very similar and $Q>0$.}
\label{fig:overlap}
\end{figure}

For supercooled liquids, the overlap quantifies the degree of similarity between the density profiles measured in two independent configurations. As sketched in Fig.~\ref{fig:overlap}, for two copies in which the density profiles are uncorrelated, $Q \simeq 0$ but the overlap becomes close to 1 for correlated configurations. The order parameter $Q$ can also be interpreted as a distance between the two copies in configuration space~\cite{parisi1983order}.

\begin{figure*}
\includegraphics[width=0.9\linewidth]{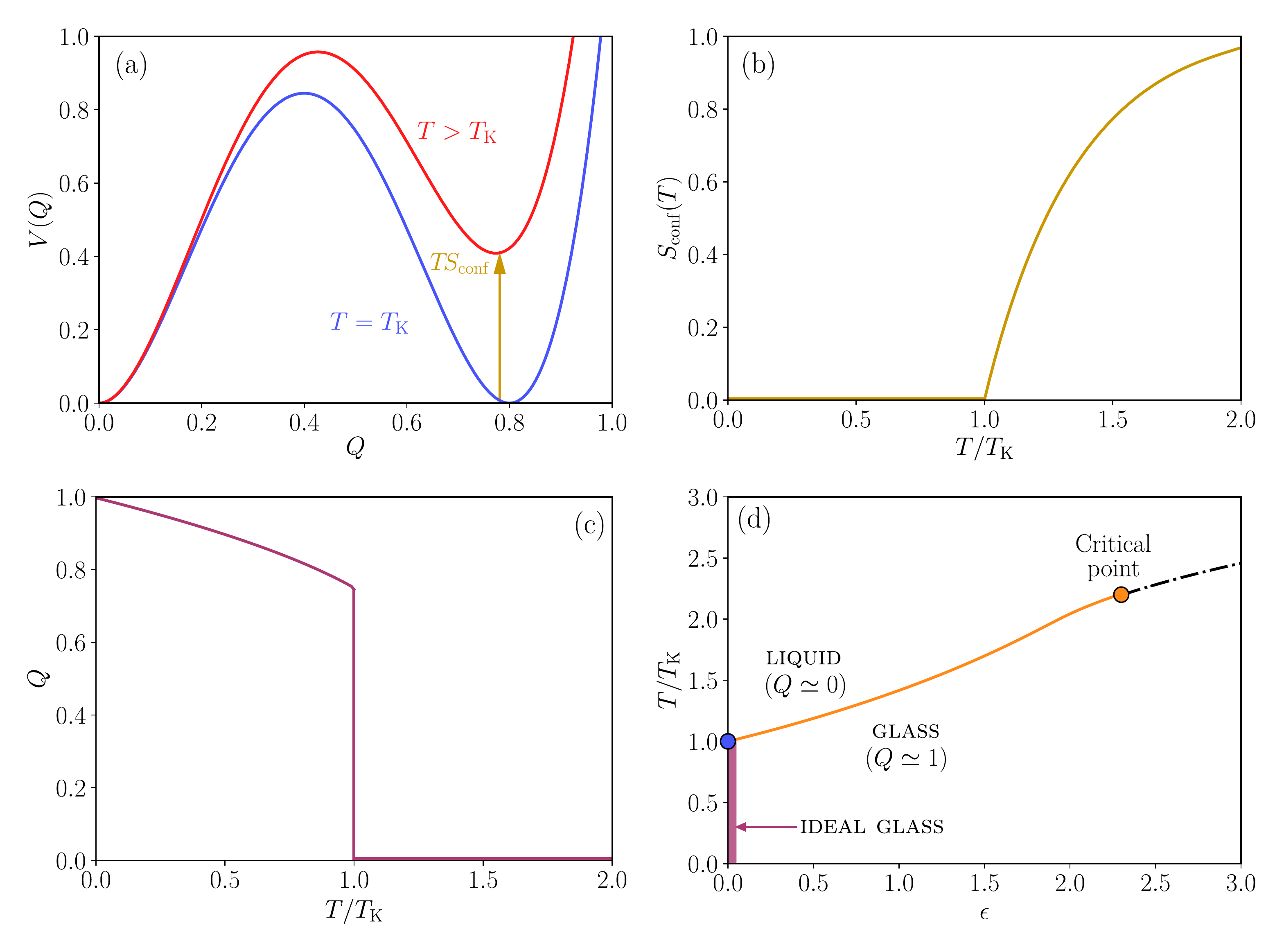}
\caption{{\bf Mean-field theory of glass formation.} (a) The Landau free energy $V(Q)$ develops a secondary minimum corresponding to the metastable glass phase approaching $T_\mathrm{K}$ from above. (b) The free energy difference between the stable liquid and the metastable glass defines the configurational entropy $S_\mathrm{conf}(T)$, which vanishes continuously at $T_\mathrm{K}$, (c) where the overlap order parameter $Q$ jumps discontinuously to a finite value. (d) Increasing the thermodynamic field $\epsilon$ conjugate to $Q$ induces a first-order transition line for $T>T_\mathrm{K}$ ending at a critical point in the universality class of the random-field Ising model. Above the critical point, the line of maximum overlap fluctuations is marked as dash-dotted line.}
\label{fig:MF}
\end{figure*}

In practice the calculation of $Q$ requires the introduction of a microscopic length scale $a$ corresponding to the amplitude of the thermal vibrations around the average positions of the particles, roughly corresponding to the cage size~\cite{guiselin2020overlap}. If one considers two copies of the liquid $\mathcal{C}_\alpha=\{\bm{r}_i^{(\alpha)}\}_{i=1\dots N}$ defined from the particle positions $\bm{r}_i^{(\alpha)}$ for $\alpha = 1$, 2, the microscopic density profiles are defined as $\rho^{(\alpha)}(\bm{r})=\sum_{i=1}^N \delta(\bm{r}-\bm{r}_i^{(\alpha)})$ with $\delta(x)$ the Dirac distribution. The overlap then reads
\begin{equation}
\begin{aligned}
Q[\mathcal{C}_1,\mathcal{C}_2]&=\frac{1}{N}\iint\mathrm{d}\bm{r}\mathrm{d}\bm{r}'\rho^{(1)}(\bm{r})\rho^{(2)}(\bm{r}')w(\vert\bm{r}-\bm{r}'\vert)\\
&=\frac{1}{N}\sum_{i,j=1}^N w(\vert\bm{r}_i^{(1)}-\bm{r}_j^{(2)}\vert),
\end{aligned}
\label{eq:overlap}
\end{equation}
with $w(x)=1$ if $x\leq a$ and 0 otherwise. From Eq.~(\ref{eq:overlap}) it is clear that $Q[\mathcal{C}_1,\mathcal{C}_2] \simeq 0$ for uncorrelated configurations, while $Q[\mathcal{C}_1,\mathcal{C}_2] \simeq 1$ for strongly correlated ones. Just as for spin glasses, it is an intensive, normalised function which quantifies the degree of similarity of the main microscopic degrees of freedom in two independent configurations. 

The calculation of the Landau free energy $V(Q)$ proceeds in two separate steps. First, one chooses and fixes a reference configuration $\mathcal{C}_0$ for the system at thermal equilibrium. The overlap between $\mathcal{C}_0$ and any other configuration $\mathcal{C}$ is $Q[\mathcal{C},\mathcal{C}_0]$, defined by Eq.~(\ref{eq:overlap}). The equilibrium probability distribution function $P(Q,\mathcal{C}_0)$ of this overlap is obtained by performing a canonical average
\begin{equation}
  P(Q,\mathcal{C}_0) = \sum_{\mathcal{C}}p_\mathrm{eq}(\mathcal{C})\delta\left(Q-Q[\mathcal{C},\mathcal{C}_0]\right),
  \label{eq:pq}
\end{equation}
where the statistical weight of any configuration $\mathcal{C}$ is controlled by the Boltzmann factor at temperature $T$ ($k_\mathrm{B} T = 1/\beta$),
\begin{equation}
  p_{\rm eq}(\mathcal{C}) = \frac{e^{-\beta\mathcal{H}[\mathcal{C}]}}{\sum_{\mathcal{C}'}e^{-\beta\mathcal{H}[\mathcal{C}']}}.
\end{equation}
By definition, the free energy $V(Q,\mathcal{C}_0)$ is the thermodynamic potential controlling the fluctuations of the overlap $Q$ for a fixed reference configuration~\cite{landau2013statistical}, 
\begin{equation}
  P(Q,\mathcal{C}_0) \sim \exp ( - \beta N V(Q,\mathcal{C}_0) ),
\end{equation}
so that the most probable value of the overlap indeed corresponds to the minimum of the free energy in the thermodynamic limit, $N \to \infty$. In a second 
step one needs to perform a final average over all possible choices for the reference configuration to get the desired averaged free energy 
\begin{equation}
V(Q) = \sum_{\mathcal{C}_0} p_{\rm eq}(\mathcal{C}_0) V(Q,\mathcal{C}_0),
\label{eq:vq}
\end{equation}
where the statistical weight of a given reference configuration is again controlled by the Boltzmann factor. The amorphous structure of the glass directly enters the canonical average in Eq.~(\ref{eq:pq}) which contains quenched disorder (the reference configuration $\mathcal{C}_0$) requiring a separate average in Eq.~(\ref{eq:vq}). It is this double averaging procedure which makes the analytic calculation of $V(Q)$ a difficult mathematical problem, even in 
the limit of infinite dimensions~\cite{parisi2020theory}.  

The temperature evolution of $V(Q)$ in the mean-field limit is shown in Fig.~\ref{fig:MF}(a). At sufficiently high temperatures, the free energy has a unique 
minimum for $Q=0$. This equilibrium state corresponds to the liquid phase in which there is a large number of distinct accessible configurations, so that two 
randomly chosen configurations are very different. At lower temperature, the free energy develops a secondary minimum for $Q>0$, which therefore represents a metastable phase. This corresponds to a phase with strong correlations between equilibrium configurations which are all amorphous. When the temperature 
decreases further, the secondary minimum becomes more and more stable. When the Kauzmann temperature $T_\mathrm{K}$ is reached, the two minima 
have the same free energy and an equilibrium phase transition takes place between the liquid with $Q=0$ and the ideal glass at $Q > 0$. This is a first-order 
phase transition because the overlap $Q$ jumps discontinuously at $T_\mathrm{K}$, as shown in Fig.~\ref{fig:MF}(c). The central conclusion for our discussion 
is that, in the mean-field limit, an equilibrium glass exists as a genuine state of matter for $T<T_\mathrm{K}$.

The discontinuity of the order parameter $Q$ at the transition indicates that equilibrium configurations in the ideal glass phase are strongly correlated. More fundamentally, the transition reflects a spontaneous symmetry breaking, which is that of the underlying replica symmetry, and the discontinuity is associated with a specific breaking scheme of this symmetry~\cite{mezard1987spin}. The liquid is symmetric with respect to the replica symmetry, because if several replicas (copies) are sampled in the liquid, no density field is privileged. Conversely, the ideal glass breaks this symmetry, as now all replicas possess a similar amorphous density profile. Physically, the ideal glass is a disordered state of matter in real space (the distribution of the average positions of the particles is amorphous), but ordered in configuration space (all equilibrium configurations look similar).

For $T>T_\mathrm{K}$, the free energy difference $\Delta V(T)$ between the two minima of $V(Q)$ represents the thermodynamic cost to force the system to remain localised near a given configuration in order to maintain a high value of $Q$, preventing the system to explore the much larger configuration space available in the liquid. The free energy cost is therefore entropic in nature. This free energy difference $\Delta V(T)$ in fact represents a natural definition of the configurational entropy $S_\mathrm{conf}(T) \equiv \Delta V(T)/T$ with the consequence that it vanishes continuously when $T\to T_\mathrm{K}$ as shown in Fig.~\ref{fig:MF}(b). This last result implies that the liquid-glass transition has no latent heat as the latent heat is normally proportional to the entropy jump at a first-order phase transition. The ideal glass transition is thus an unconventional example of a discontinuous transition~\cite{kirkpatrick1989scaling}. More strikingly perhaps, the continuous vanishing of the configurational entropy in Fig.~\ref{fig:MF}(b) exactly realises the scenario in which the ideal glass transition corresponds to the entropy crisis first put forward by Kauzmann~\cite{kauzmann1948nature}. With hindsight, Kauzmann's work appears truly visionary.

This exposition of the mean-field theory of the glass transition closely follows traditional descriptions  of the van der Waals model for the liquid-gas transition or the Curie-Weiss model to describe ferromagnetism. Pushing these analogies further, one can introduce the field $\epsilon$ that is thermodynamically coupled to the order parameter $Q$~\cite{franz1997phase}. This field plays a role similar to the magnetic field $H$ conjugate to the magnetisation $M$ in the Curie-Weiss model. It introduces a new contribution $- \epsilon Q$ in the Hamiltonian which for $\epsilon>0$ makes large $Q$ values energetically more favourable. In the presence of a field, equilibrium is found by looking for the minimum of the tilted free energy $V(Q)-\epsilon Q$ for fixed ($T$,~$\epsilon$). When the glass is metastable for $T>T_\mathrm{K}$, increasing $\epsilon$ triggers a first-order phase transition towards a high-overlap state, see Fig.~\ref{fig:MF}(d). Physically, this can be understood from an energy-entropy argument, as for many equilibrium phase transitions~\cite{goldenfeld2018lectures}. When $\epsilon$ is increased, the system gains energy by adopting a large value of $Q$ but loses configurational entropy because it only visits a set of highly-correlated configurations. Beyond a threshold value $\epsilon^*(T)\simeq TS_\mathrm{conf}(T)=\Delta V(T)$, the system ends up in a high-overlap phase. As a result, the existence of a discontinuous phase transition at the Kauzmann temperature $T_\mathrm{K}$ implies the existence of a first-order transition line in the ($T$,~$\epsilon$) plane. This line ends at a second-order critical point, once again in full analogy with the liquid-gas transition in the van der Waals description. 

The behaviour of the order parameter in the vicinity of a critical point defines universality classes. Each universality class is characterised by a different set of critical exponents. For glasses, the presence of disorder in the various canonical averages in Eqs.~(\ref{eq:pq}, \ref{eq:vq}) implies that the critical end-point at a finite applied field $\epsilon$ belongs to the universality class of the random-field Ising model~\cite{biroli2014random, franz2013universality}. (See Ref.~[\onlinecite{nattermann1998theory}] for an introduction to the random-field Ising model.) By contrast, the critical points of the liquid-gas and paramagnetic-ferromagnetic phase transitions both belong to the universality class of the Ising model with no disorder. The transition between a paramagnet and a spin glass is also controlled by the presence of quenched disorder but it forms a universality class on its own~\cite{mezard1987spin}. Despite the mathematical similarities between molecular glasses and conventional spin glasses, in particular the definition of the order parameter, the emergence of a glassy phase is qualitatively distinct in these two systems.

Before closing this section, it is instructive to pause and ask: should one be satisfied that after decades of theoretical work on the problem of the glass transition, the only solid set of results is about the mean-field theory? There are several answers. First, the study of phase transitions and critical phenomena in the presence of quenched disorder is notoriously more complicated than for pure, conventional systems~\cite{de2006random}. It should thus come as no surprise that the statistical mechanics of glasses is a very difficult analytic problem, and the existence of a solid starting point is an important step. However, past experience with disordered systems also suggests that including corrections to mean-field is going to be an even harder problem. At present it is not known whether and how the mean-field construction can survive the inclusion of fluctuations that are neglected in the mean-field theory. It is conceivable that the very existence of an ideal glass as a phase of matter, established within the mean-field approximation, becomes impossible when the relevant fluctuations are understood and included in the calculations.    

\section{Does the ideal glass exist beyond mean-field?}

\label{sec:numerics}

The ideal glass, which is defined in full equilibrium conditions, differs from the conventional material prepared in the laboratory that has fallen out of equilibrium near $T_\mathrm{g}$ while cooling. How can one observe an ideal glass? The most direct way to detect the ideal glass consists in the quasi-static cooling of two copies of the supercooled liquid to maintain thermal equilibrium. A discontinuous jump of their relative overlap $Q$ would be observed at the Kauzmann temperature $T_\mathrm{K}$. This is conceptually very simple! This seems however impossible due to the dramatic slowing down of the dynamics of the supercooled liquid at low temperature which prevents an ergodic exploration of the configuration space below $T_\mathrm{g}$.

The difference between ideal and conventional glasses appears subtle: the conventional glass is confined in configuration space because it moves too slowly to explore the many different available states, while the ideal glass is confined in configuration space because there is nowhere to go even given an infinite amount of time. Establishing this difference for temperatures where equilibration is extremely long is then obviously a challenge. In addition, the first-order nature of the equilibrium liquid-glass transition implies that it cannot be simply revealed by the growth of some critical fluctuations, as would be the case when approaching a second-order phase transition. More subtle signs of an incipient phase transition need to be detected.

Major progress has recently been made to tackle these two distinct problems in computer simulation studies. To solve the first problem, efficient Monte Carlo algorithms have been developed which are able to easily sample the configuration space in equilibrium conditions employing non-physical particle motions~\cite{grigera2001fast, berthier2016equilibrium, ninarello2017models, Berthier_2019, parmar2020ultrastable}. As a result, equilibrium thermodynamic measurements can be performed at low temperatures where this task would be impossible by simply simulating the physical dynamics. In essence, these algorithms solve the first problem and maintain equilibrium at temperatures much lower than in conventional simulations. For certain numerical models, it is estimated that temperatures smaller than the experimental glass temperature $T_\mathrm{g}$ can be equilibrated, thus provinding access to the experimental temperature gap between $T_\mathrm{g}$ and $T_\mathrm{K}$.

\begin{figure*}
\includegraphics[width=0.9\linewidth]{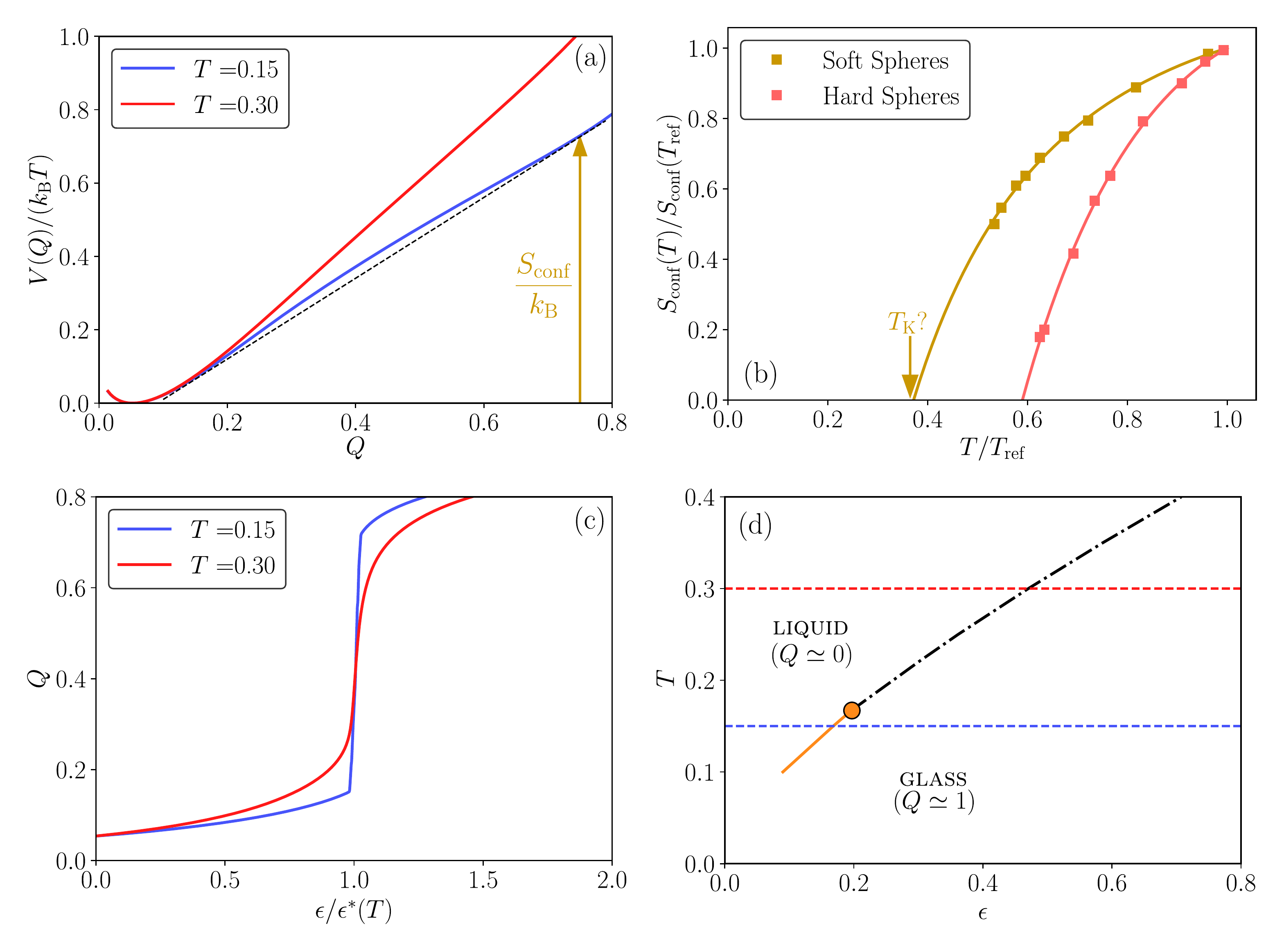}
\caption{{\bf Numerical results of three-dimensional systems reveal features of the mean-field theory.} (a) Temperature evolution of Landau free energy $V(Q)$ which develops a singularity at large $Q$ consistent with the existence of a metastable glass phase. The dashed line highlights the non-convexity of the free energy at the lowest temperature. (b) Configurational entropy measured in two glass-forming models. The free energy difference between the equilibrium liquid and metastable glass decreases rapidly with decreasing $T$ and seems to extrapolate to a finite Kauzmann temperature $T_\mathrm{K}$. (c) The growth of the overlap upon application of a field $\epsilon$ is smooth at large $T$ but becomes discontinuous at low $T$, revealing a first-order equilibrium transition line $\epsilon^*(T)$. (d) Equilibrium phase diagram ($T$,~$\epsilon$). Dashed lines are temperatures shown in (a, c). The dash-dotted line represents the location of the inflection point of $Q(\epsilon)$ which transforms in a discontinuity (full line) below a second-order critical point (dot).}
\label{fig:numerics}
\end{figure*}

The second problem concerns numerical signatures of the approach to a first-order phase transition. The most direct measurement amounts to directly estimating 
the Landau-Franz-Parisi thermodynamic potential $V(Q)$ across a range of temperatures. This program was started long ago~\cite{franz1998effective,cardenas1999constrained} but was plagued by both equilibration problems and sampling issues. More recently, improved 
methods to assess $V(Q)$ from Eq.~(\ref{eq:pq}) based on large deviations techniques have been proposed~\cite{berthier2013overlap} and combined 
with enhanced sampling techniques~\cite{berthier2014novel}. As a result, the Landau free energy $V(Q)$ has been measured in several numerical models of 
glass-forming liquids in dimension $d=2$ and $d=3$ down to very low temperatures~\cite{berthier2013overlap, berthier2015evidence, berthier2017configurational, guiselin2022statistical}. The expected behaviour for $V(Q)$ has been confirmed directly in these computer simulations, as illustrated in Fig.~\ref{fig:numerics}(a), 
with the emergence of singularities at large $Q$ when $T$ is low enough. Since these measurements are performed in equilibrium conditions and do not 
require any extrapolation, they are direct proof that the system is on its way to an equilibrium first-order phase transition with an emerging metastable glass 
phase at large $Q$.  

Glass metastability can be established even more clearly by coupling the overlap $Q$ to its conjugate field $\epsilon$~\cite{berthier2013overlap, berthier2015evidence, guiselin2020random, guiselin2022statistical, guiselin2022static}. In three dimensions, one finds a discontinuous jump of the overlap as $\epsilon$ is increased at constant $T$ beyond a critical value $\epsilon^*(T)$, see Fig.~\ref{fig:numerics}(c). This discontinuous jump disappears when the temperature is too large, as expected from mean-field theory. 

As a result, a line of first-order transition $\epsilon^*(T)$ can be determined in the phase diagram ($T$,~$\epsilon$), see Fig.~\ref{fig:numerics}(d), which ends at a critical point whose universality class was recently analysed by detailed finite-size scaling analysis~\cite{guiselin2020random, jack2016phase}. (At the critical point, the amplitude of the order parameter fluctuations only diverges in the thermodynamic limit, so that analysis of this divergence with system size yields information on the critical exponents~\cite{newman1999monte}.) Results in three-dimensional systems appear compatible with the predicted random-field Ising model universality class. 

Interestingly, the connection with the random-field Ising model implies that neither the critical point nor the first-order transition line can exist in two dimensions at any finite temperature. This is because the lower critical dimension~\cite{goldenfeld2018lectures} of the random-field Ising model is $d=2$, so that for $d \leq 2$ there is no longer a finite transition temperature in this model~\cite{imry1975random}. This non-trivial prediction was recently confirmed in computer simulations of a two-dimensional glass-forming model~\cite{guiselin2022statistical}.

Using similar techniques~\cite{berthier2019configurational}, the configurational entropy $S_\mathrm{conf}(T)$ was recently estimated in computer models down 
to extremely low temperatures, including below the experimental $T_\mathrm{g}$, see Fig.~\ref{fig:numerics}(b). These measurements on the one hand confirm 
and amplify the results obtained from calorimetric experiments, such as the results discussed in Kauzmann's original work~\cite{kauzmann1948nature}. Whereas 
the nature of the configurational entropy obtained experimentally is subject to discussions~\cite{tatsumi2012thermodynamic}, in the simulations $S_{\rm conf}(T)$ 
can be interpreted as the free energy difference between the equilibrium liquid and metastable glass. The steep decrease observed in Fig.~\ref{fig:numerics}(b) 
at low temperatures shows that the glass could become the equilibrium phase below a finite temperature $T_\mathrm{K}$. Two-dimensional simulations~\cite{berthier2019zero} are compatible with a value $T_\mathrm{K}=0$, again suggesting that $d=2$ plays a special role with a zero-temperature critical point and no equilibrium glass phase at any $T>0$.

Finally, metastability near a first-order phase transition is generically associated with an important length scale related to the physics of nucleation. If a system is prepared in a metastable phase, the emergence of the stable phase proceeds by first nucleating a critical droplet of the stable phase which can then invade the entire system. The minimal size of the nucleating droplet is inversely proportional to the free energy difference between the two phases and diverges at the transition. This behaviour is generic to first-order phase transitions and describes for instance how crystals form below $T_\mathrm{m}$. In the present case, one would essentially like to estimate the size of the critical nucleus of the stable liquid from the metastable glass above $T_\mathrm{K}$. However, considering the unconventional first-order nature of the transition at $T_\mathrm{K}$, this is far from trivial and special computational tools are needed. Such tools have been proposed~\cite{bouchaud2004adam} and implemented~\cite{biroli2008thermodynamic, berthier2016efficient, yaida2016point, berthier2017configurational}, allowing one to estimate this important length scale, $\xi(T)$, which is also known as the point-to-set length scale~\cite{montanari2006rigorous}.  

Numerical measurements show that $\xi(T)$ increases when the temperature $T$ decreases, as expected when approaching a first-order phase transition. The temperature dependence of $\xi(T)$ is compatible with the above nucleation argument, which suggests that $\xi(T) \sim 1/S_{\rm conf}(T)$. The growth of the point-to-set length scale $\xi(T)$ therefore accompanies the steep decrease in the configurational entropy which signals that glass and liquid free energies are getting closer as $T$ decreases towards $T_\mathrm{K}$.  

\section{Conclusion} 

\label{sec:discussion}

State-of-the-art computer studies provide solid confirmations that supercooled liquids carefully maintained in equilibrium conditions at low temperatures (even below $T_\mathrm{g}$) display thermodynamic fluctuations that are fully compatible with the mean-field description of an incipient discontinuous 
liquid-glass phase transition. These observations invalidate the well-known motto that that glass formation is essentially a dynamic process accompanied by little structural change. Of course, the existence of the transition at a finite Kauzmann temperature $T_\mathrm{K}>0$ remains to be demonstrated by direct measurements. This would represent a major objective for future work. From a computational viewpoint, this requires the development of numerical algorithms more sophisticated than current ones in order to achieve equilibration even closer to the putative $T_\mathrm{K}$.

What about experiments? Seventy years after Kauzmann~\cite{kauzmann1948nature}, calorimetric measurements provide estimates of a configurational entropy 
which robustly extrapolate to a finite Kauzmann temperature~\cite{tatsumi2012thermodynamic}. In recent years, indirect experimental estimates of the point-to-set correlation length scale $\xi(T)$ were also obtained via non-linear dielectric susceptibilities~\cite{albert2016fifth, biroli2021amorphous}. (If the frequency of the 
imposed oscillatory electric field is larger than the typical relaxation frequency of the supercooled liquid, then the liquid approximately behaves like a mosaic of independent glasses of typical size $\xi(T)$~\cite{kirkpatrick1989scaling}, each one of them responding with an induced dipolar moment that depends on their 
typical size $\xi(T)$, which can then be inferred.)

The seemingly impossible task to achieve equilibration below $T_\mathrm{g}$ was overcome after the discovery that amorphous glassy films prepared by physical vapour deposition are much closer to equilibrium than their liquid-cooled counterparts~\cite{swallen2007organic}. When the deposition rate is sufficiently low 
(of the order of 1~nm/s), and the temperature judiciously chosen (about 15~\% lower than the experimental glass transition temperature $T_\mathrm{g}$), it 
has become possible to create glassy films that seem to remain in equilibrium much closer to $T_\mathrm{K}$ than in any previous work~\cite{beasley2019vapor}, 
thus providing novel support for previous extrapolations of the configurational entropy from above $T_\mathrm{g}$. These new glassy materials are ``ultrastable'', because they correspond to thermodynamic conditions impossible to achieve using bulk cooling~\cite{ediger2017perspective}.

In addition to providing equilibrium materials much closer to $T_\mathrm{K}$, the transformation kinetics of an ultrastable glass into a liquid upon heating closely resembles the nucleation and growth process commonly observed near a first-order phase transition~\cite{kearns2010one,jack2016melting}. For thin ultrastable films, the transformation is initiated at the free surface and propagates as a front. This is reminiscent of heterogeneous nucleation. For thicker samples, liquid droplets first nucleate in the bulk and their size then increases until complete transformation, reminiscent of homogeneous nucleation~\cite{vila2020nucleation}. These analogies reinforce the idea that ultrastable glasses behave very much like a distinct phase of matter accessed via a discontinuous phase transition. Computer simulations of this transformation process~\cite{fullerton2017density, flenner2019front} could help unravel the main mechanisms at play, and could allow direct quantitive comparisons with predictions from classical nucleation theory~\cite{Kalikmanov2013}.

The construction of a complete theoretical description of the nature of the glass remains an active research area. The progress summarised here results from a synergy between mathematical work, experimental discoveries and the development of novel computational tools. This quest already gave rise to many fruitful developments, as recognized by the 2021 Nobel Prize in Physics awarded to Giorgio Parisi who %was, and still is, an active player 
has been a driving force in this field. In addition to understanding glass more deeply, tools and concepts developed for glass physics are at the core of deep analogies between a multitude of complex systems, such as neural networks in biology~\cite{mezard1987spin, amit1985spin}, algorithms for constrained optimisation~\cite{antenucci2019glassy}, or, more recently, machine learning~\cite{zdeborova2016statistical, bahri2020statistical}.

\begin{acknowledgments}
This work was supported by a grant from the Simons Foundation (Grant No. 454933, L.B.).
\end{acknowledgments}

\bibliographystyle{apsrev4-1}

\bibliography{biblio.bib}

\end{document}